\documentclass[twocolumn,showpacs,preprintnumbers,amsmath,amssymb]{revtex4}

\usepackage{graphicx}
\usepackage{dcolumn}
\usepackage{bm}

\begin{document}

\preprint{APS/123-QED}

\title{Omni-directional, broadband and polarization-insensitive thin absorber in the terahertz regime}

\author{Yuqian Ye$^{1,3}$ }
\author{Yi Jin$^1$}
\author{Sailing He$^{1,2,}$}
 \email{sailing@kth.se}
\affiliation{%
$1.$Centre for Optical and Electromagnetic Research, State Key
Laboratory of Modern Optical Instrumentation,
Zijingang Campus, Zhejiang University, China\\
$2.$ Division of Electromagnetic Engineering, School of Electrical
Engineering, Royal Institute of Technology, S-100 44 Stockholm,
Sweden\\
$3.$Department of Physics, Zhejiang University, Hangzhou 310027,
China
}%

\date{\today}

\begin{abstract}
A nearly omni-directional THz absorber for both transverse electric
(TE) and transverse magnetic (TM) polarizations is proposed. Through
the excitation of magnetic polariton in a metal-dielectric layer,
the incident light is perfectly absorbed in a thin thickness which
is about $25$ times smaller than the resonance wavelength. By simply
stacking several such structural layers with different geometrical
dimensions, the bandwidth of this strong absorption can be
effectively enhanced due to the hybridization of magnetic polaritons
in different layers.
\end{abstract}

\maketitle

\section{Introduction}
High absorption in a thin film is of critical importance in some
device applications such as micro bolometers \cite{Mauskopf},
thermal detectors \cite{Parsons} and solar cells \cite{Rand,Pillai},
whose working spectra are from infrared to optical frequencies.
Several absorbing structures have been proposed previously. For
instance, a bare metallic grating can fully absorb the incident
light at some well defined wavelengths
\cite{Hutley,Collin,Tan,Popov}. However, absorption in metallic
gratings usually relies on delocalized surface excitations
\cite{Bliokh}, which are highly sensitive to the angle of incidence.
This narrow-angle absorption prevents their application to e.g.
photovoltaic cells, where wide collection angles are necessary.
Another widely used method is based on structured metal-dielectric
interfaces, such as Dallenbach layers \cite{Reinert, Knott} and
Salisbury screens \cite{Salisbury,Engheta}, which can give
frequency-selective absorption of incident light. However, these
types of structures often limited to a minimal thickness of one
quarter wavelength. Recently, Landy et al. proposed an innovative
metal-dielectric composite named metamaterial absorber to overcome
this thickness limitation \cite{Landy1,Tao1}. Later, some efforts
have been made on this metamaterial absorber to achieve
polarization-insensitive absorption \cite{Landy2} or wide-angle
(only in the incident plane with a fixed azimuthal angle) absorption
\cite{Tao2,Avitzour}. The above-mentioned absorbing structures have
shown their great ability for light absorption. However,
omni-directional absorption (i.e., wide-angle absorption for any
azimuthal angle) in a thin film (much thinner than one quarter
wavelength) for incident wave of both polarizations remains a
challenge. Moreover, most of these designs are based on strong
electromagnetic resonances to effectively absorb the incident light
in a thin thickness, especially for a metamaterial absorber, and
consequently the bandwidth of this resonant absorption is often
narrow, typically no more than $10\%$ with respect to the center
frequency. This narrow bandwidth feature of the resonant absorption
limits the device applications of these absorbing structures.

This paper is dedicated to the design of an omni-directional and
broadband thin absorber for both polarizations in terahertz
frequency range ($0.1-10$ ${\rm THz}$), where it is difficult to
find naturally occurring materials with very strong absorption
\cite{Williams,Tonouchi}. We numerically demonstrate that a
composite structure of a cut-wire array, a lossy polymer separation
layer and a metal ground film can be used as an effective absorber
with a resonant absorption up to $99.9\%$ when the thickness of the
separation layer is properly designed. Moreover, by simply replacing
the cut-wires with crosses, nearly omni-directional absorption is
achieved in this more symmetric composite structure for both TE and
TM polarizations. Most importantly, we successfully demonstrate
that, by stacking these layers of metallic crosses with different
geometrical dimensions, several closely positioned resonant peaks
are merged in the absorption spectrum due to the hybridization of
the magnetic polaritons, and consequently the bandwidth is
effectively enhanced.

\section{Wide angle absorption of a composite cut-wire structure}

\begin{figure}
\includegraphics[width=3.5in]{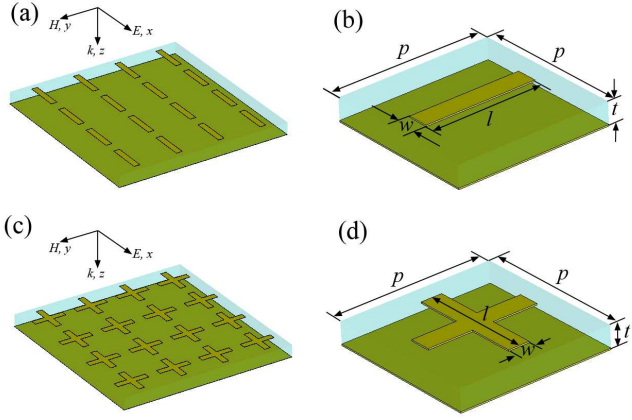}
\caption{\label{Fig1} Schematic diagrams of the THz absorbers
consisting of a metallic film, a polymer separation layer, and (a) a
cut-wire array; (c) a cross array. (b) and (d) show the unit cell of
the two absorbers in the calculation of the reflection spectrum.
Axes in (a) and (c) indicate the polarization and propagation
direction of the incident wave. }
\end{figure}

\begin{figure}
\includegraphics[width=3.5in]{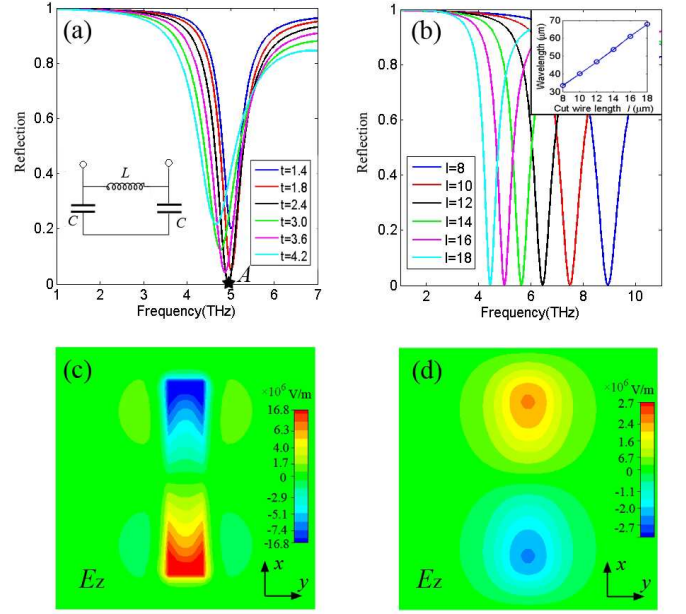}
\caption{\label{Fig2} (color online) Reflection spectra for (a)
different values of polymer separation thickness $t$; (b) different
values of cut-wire length $l$. Star point \textbf{A} denotes the
resonant dip when $t = 2.4$ ${\rm {\mu}m}$. (c) and (d) show the
distributions of the z-component electric field for resonance
\textbf{A} on the cut-wire and the metallic film, respectively. The
inset of Fig. $\ref{Fig2}$(a) shows the effective LC circuit for the
magnetic polariton resonance. The inset of Fig. $\ref{Fig2}$(b)
shows the resonant wavelength as a function of the cut-wire length.
}
\end{figure}

First we study a simplified composite structure which is made of a
metallic cut-wire array positioned above a metallic film (with a
polymer layer in between) as depicted in Fig. $\ref{Fig1}$(a). A
lossy polymer with dielectric $\epsilon = 3.5+0.2i$ is used in our
numerical simulation. The thickness $t$ of the polymer layer is
adjustable, the width $w$ and length $l$ of the cut-wire are $3$
${\rm {\mu}m}$ and $16$ ${\rm {\mu}m}$, respectively. As shown in
Fig. $\ref{Fig1}$(b), the length of the unit cell (i.e., the period
of the cut-wire array) in both x and y directions are $p = 22$ ${\rm
{\mu}m}$. In the simulation, the metallic structures were made of
lossy gold with a conductivity of $\sigma = 4.09\times10^7$ ${\rm
S/m}$ and a thickness of $200$ ${\rm nm}$. As the thickness of the
metallic film used here is much larger than the typical skin depth
in the THz regime (to avoid transmission through the metallic film),
the reflection is the only factor limiting the absorption. Here we
consider the case when a plane wave [with electric field polarized
in x direction; see Fig. $\ref{Fig1}$(a)] normally impinges on the
structure. By using a Finite-Integration Time Domain algorithm
\cite{CST}, the reflection spectra for different separation
distances $t$ are calculated and shown in Fig. $\ref{Fig2}$(a). In
each reflection spectrum shown in Fig. $\ref{Fig2}$(a), one sees a
significant resonant dip where the reflection drops rapidly. To
obtain a physical insight, the distributions of the z-component
electric field on the cut-wire and the metallic film are shown in
Fig. $\ref{Fig2}$(c) and (d), respectively, for resonance \textbf{A}
[see Fig. $\ref{Fig2}$(a)] as an example. As shown in Fig.
$\ref{Fig2}$(c), charges of opposite signs accumulate at the two
ends of the cut-wire, indicating the excitation of an electric
dipole resonance on the cut-wire. This electric dipole is greatly
coupled with its own image, which oscillates in anti-phase on the
metallic film [see Fig. $\ref{Fig2}$(d)]. Consequently, a magnetic
polariton (or "magnetic atom") \cite{liu1,Li} is formed, which
induces a strong magnetic response (see Fig. $\ref{Fig6}$(a) below)
and causes a resonant dip in the reflection spectrum \cite{liu2}.
The coupling strength of the electric dipoles as well as the
magnetic response is mainly determined by the separation distance.
By tuning the thickness of the spacer layer, we can obtain an
optimal value $t = 2.4$ ${\rm {\mu}m}$, at which the electric and
magnetic response makes the composite structure impedance-matched to
the free space, and the reflection is considerably suppressed ($<
0.1\%$) at resonant frequency $\omega = 4.96$ ${\rm THz}$, as shown
by the black curve in Fig. $\ref{Fig2}$(a). $99.9\%$ of the incident
wave is effectively absorbed in the composite structure with a
thickness of less than $\lambda/25$. To give a further
interpretation, an LC-circuit model, as shown in the inset of Fig.
$\ref{Fig2}$(a), is introduced to approximately describe this
magnetic polariton caused resonance. Each of the two capacities is
formed by the metallic film and an upper or lower half of the
cut-wire. The capacitance $C$ can be described approximately by a
two-plate capacitor formula $C \sim (w¡¤l /2)/t$ \cite{Zhou}. The
inductance L of the structure is approximately given by $L \sim
(l¡¤t)/w$ (as for the case of two parallel plates). Then the
resonant frequency is given by:

\begin{equation}
f_m=\frac{1}{2\pi\sqrt{LC/2}}\sim \frac{1}{l} \label{Eq1}
\end{equation}

From Eq. (1), we can deduce that the resonant frequency is not
sensitive to the change of separation distance $t$. This agrees well
with our simulation results (the resonant peaks are at similar
frequencies), as illustrated in Fig. $\ref{Fig2}$(a). The reflection
spectra for different lengths of the cut-wire are also shown in Fig.
$\ref{Fig2}$(b). Compared with Fig. $\ref{Fig2}$(a), a significant
redshift of the resonant peak (with little change in amplitude) is
observed when $l$ increases from $8$ ${\rm {\mu}m}$ to $18$ ${\rm
{\mu}m}$. The linear dependence of the resonant wavelength on
cut-wire length $l$ is demonstrated in the inset of Fig.
$\ref{Fig2}$(b), as predicted by Eq. (1).

\begin{figure}
\includegraphics[width=3.5in]{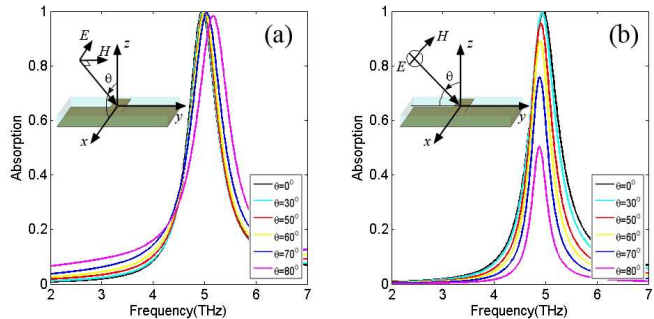}
\caption{\label{Fig3} (color online) Absorption spectra for
different incidence-angles with (a) TM and (b) TE configurations.
The insets depict the polarization and propagation direction of
incident waves.  }
\end{figure}

Next we study the resonant absorption behavior for the case of
optimal thickness $t = 2.4$ ${\rm {\mu}m}$, when the incident angle
increases. Figs. 3(a) and 3(b) give the absorption spectra when
incident angle $\theta$ varies from $0$ to $80^\circ$ in TM
configuration (i.e., H-field is fixed along the y direction) and TE
configuration (E-field is fixed along the x direction),
respectively. For the case of TM configuration, as the incident
angle increases, the amplitude of the absorption peak decreases
slightly but still keeps above $98\%$ [see Fig. $\ref{Fig3}$(a)]
even when $\theta = 80^\circ$, at which angle the electric field is
nearly normal to the cut-wire. Thus we can conclude that the
magnetic polariton, which causes the resonant absorption in the
composite structure, is effectively excited by the y-component of
the magnetic field, while the x-component of the electric field
(which can drive the electric dipole oscillation on the cut-wire)
contributes a little. Meanwhile, a blueshift of the absorption peak
occurs (see Fig. $\ref{Fig3}$(a)), which is relatively small
($<0.05$ ${\rm THz}$) for $\theta < 60^\circ$, and reach $0.2$ ${\rm
THz}$ when $\theta$ increases to $80^\circ$. It is not difficult to
understand this blueshift for the obliquely incident light. As we
know, for the normal incident case, the electric dipole of the
cut-wire [see Fig. $\ref{Fig2}$(a)] in each unit cell oscillates in
phase. Accordingly, the attractive force of the opposite charges of
the adjacent cut-wires in x direction reduces the restoring force of
the charge oscillation inside the cut-wire. Compare with a single
cut-wire, the resonant frequency of the cut-wire array is reduced
due to this interaction between the neighboring unit cells. (Our
numerical simulation has shown that this small redshift of the
absorption peak will be reduced when the period increases, i.e., the
interaction between the neighboring unit cells decreases.) However,
as the incident angle increases for the TM configuration [see the
inset of Fig. $\ref{Fig3}$(a)], the dipole oscillation of the
adjacent cut-wires in x direction is no longer in phase, which
causes the reduction of the attractive force between these adjacent
cells, and consequently the absorption peak moves slightly to a
higher frequency. For the TE configuration, a distinct amplitude
reduction of absorption peak is observed as the incident angle
increases [see Fig. $\ref{Fig3}$(b)]. However, for small incident
angle $\theta = 30^\circ$, the absorption spectrum are almost
overlapped with the one for normal incidence. A high absorption of
$89\%$ at resonant frequency is still achieved when $\theta =
60^\circ$. Beyond this angle, the amplitude of the absorption peak
drops quickly, as the y-component of the incident magnetic field
decreases rapidly to zero and can no longer efficiently excite this
magnetic polariton. Compared with the TM configuration, here the
frequency shift of the absorption peak is really small (almost
imperceptible), since the coupling of the adjacent unit cells in y
direction is relatively small due to the larger spacing. Notably,
the present simulation results demonstrate that this simplified
structure can have quite wide angle absorption for both TE and TM
configurations, which is similar to the composite structure (of
electric inductive-capacitive resonators and a metallic film) used
in Ref. \cite{Tao2}.

\section{Omni-directional and polarization-insensitive absorption in a composite structure of metallic crosses}

\begin{figure}
\includegraphics[width=3.5in]{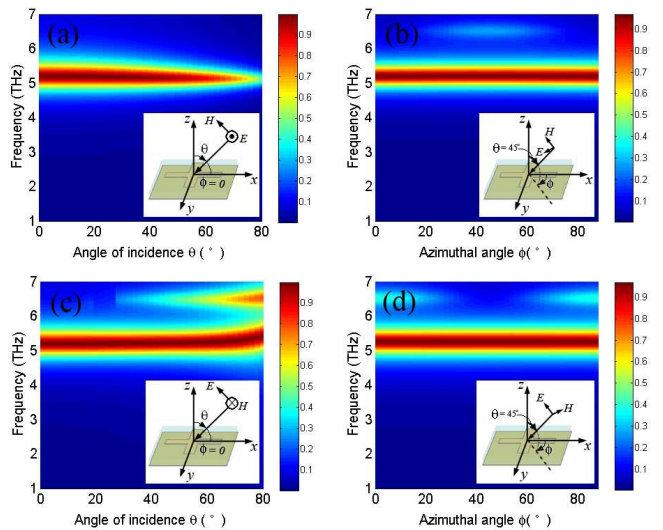}
\caption{\label{Fig4} (color online) The absorption spectra as a
function of incident angle $\theta$ for (a) TE polarization; (c) TM
polarization, when azimuthal angle $\phi = 0$. The absorption
spectra as a function of  azimuthal angle $\phi$ for (b) TE
polarization; (d) TM polarization, when incident angle $\theta =
45^\circ$. }
\end{figure}

Although the resonant absorption of the composite structure of
cut-wires can work over a wide range of incident angles $\theta$,
the effective absorption can be obtained only for a plane wave
without x-component of magnetic field. However, due to the
structural simplicity, the above absorber can be easily extended to
a more symmetric structure in order to highly absorb the plane wave
with x-component of the magnetic field. We demonstrate this by
simply replacing the cut-wire array with the cross array as shown in
Fig. $\ref{Fig1}$(c) and (d). The unit cell for the numerical
calculation is depicted in Fig. $\ref{Fig1}$(d). The cross here is
composed of two orthogonally intersected cut-wires of same size. The
width $w$ and length $l$ of the constitutive cut-wire are the same
as those used in Fig. $\ref{Fig1}$, while the optimal thickness $t$
of the separation layer is adjusted to $2.2$ ${\rm {\mu}m}$
(slightly different from the previous value) in order to be
impedance-matched to the free space. Fig. $\ref{Fig4}$(a) and (c)
show the absorption spectra as a function of the incidence-angle
$\theta$ for TE and TM waves, respectively, when the azimuthal angle
$\phi$ is fixed to 0. The propagation and the polarization direction
of the incident wave are shown in the inset of each figure. From
Fig. $\ref{Fig4}$(a) and (c) one can see that the present absorber
can still work over a wide range of incident angle $\theta$ (as for
the case of the composite cut-wire structure (see Fig.
$\ref{Fig3}$)] for both TE and TM polarizations, because the
additional constitutive cut-wire, which is normal to the electric
field and parallel to the plane of magnetic field, has no resonant
electromagnetic response, and consequently has little influence on
the excitation of magnetic polariton. Moreover, for any fixed
incident angle $\theta$, the resonant absorption of this more
symmetric structure is quite isotropic in the x-y plane (as the
azimuthal angle $\phi$ varies) as illustrated in Fig.
$\ref{Fig4}$(b) and (d) for TE and TM waves, respectively (here we
choose incident angle $\theta = 45^\circ$ as an example). The
insensitivity to the azimuthal angle variation is not difficult to
understand. Any incident plane wave impinging on this composite
structure can be decomposed into two components with the magnetic
field in x-z or y-z plane. At resonant frequency, each component can
effectively excite the corresponding magnetic polariton (between one
constitutive cut-wire and the metallic film), and then will be
greatly absorbed. Our simulation results reveal that the present
composite structure of metallic crosses gives nearly
omni-directional and polarization-insensitive absorption as the case
in Ref. \cite{Teperik}. However, compared with the absorbing
structure of nanocavities in gold substrate proposed in Ref.
\cite{Teperik}, the present structure is much thinner ($<
\lambda/25$), and can be easily manufactured with standard planar
micro-fabrication techniques \cite{Boltasseva}.

\section{Broadband absorption of a multi-layer structure}

\begin{figure}
\includegraphics[width=3.5in]{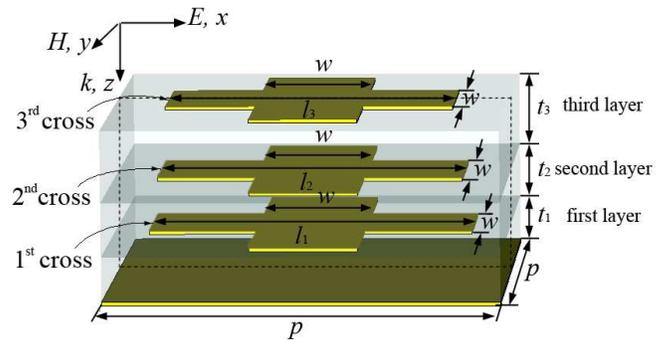}
\caption{\label{Fig5} (color online) Schematic diagram of a 3-layer
cross structure with the geometrical parameters of each layer marked
on it. The crosses from the bottom to the top are denoted in
sequence as the $1^{st}$ cross, $2^{nd}$ cross, and $3^{rd}$ cross.
The dashed line shows the cross section plane $y = 0$.  }
\end{figure}

\begin{figure}
\includegraphics[width=3.5in]{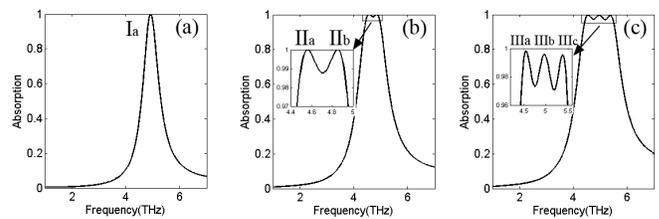}
\caption{\label{Fig6} (Absorption spectra for (a) 1-layer cross
structure; (b) 2-layer cross structure; (c) 3-layer cross structure.
The geometrical parameters used here are given in Table
$\ref{tab:table1}$. The insets show the details at the frequency
range of the resonant absorption. I$_a$, II$_a$, II$_b$, III$_a$,
III$_b$, III$_c$ denote the resonant peaks in each spectrum. }
\end{figure}

\begin{figure}
\includegraphics[width=3.5in]{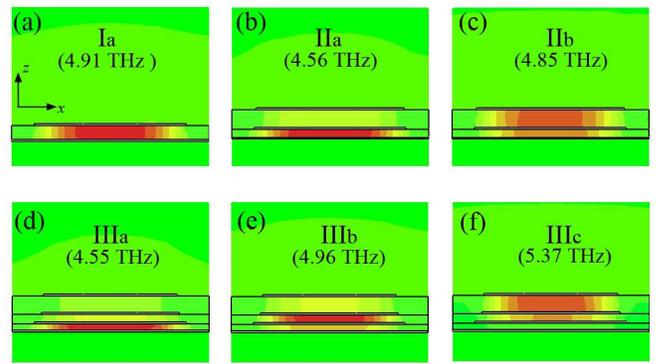}
\caption{\label{Fig7} (color online) The corresponding distributions
of y-component magnetic field magnitude $|Hy|$ in plane $y = 0$ for
each resonance shown in Fig. 6. }
\end{figure}

\begin{figure}
\includegraphics[width=3.5in]{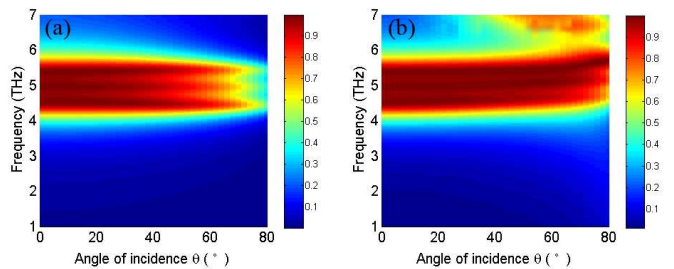}
\caption{\label{Fig8} (color online) The absorption spectra of the
3-layer cross structure as a function of the incidence-angle
$\theta$ for (a) TE polarization; (c) TM polarization, when the
azimuthal angle is fixed to $\phi = 0$. }
\end{figure}

Below we try to increase the bandwidth of the absorption by using a
multi-layer structure which can support several resonant modes
closely positioned in the absorption spectrum. From Eq. (1), the
resonant frequency of the absorption caused by the magnetic
polariton is primarily determined by wire length $l$. Thus in
different layers we design crosses of slightly different lengths to
ensure that the resonance frequencies of magnetic polaritons could
be close to each other. Then, by tuning the polymer separation
thickness of each layer, the multi-layer structure can be
impedance-matched to the free space at each resonant frequency
(similar to the one-layer case shown in Fig. $\ref{Fig2}$(a)). In
Fig. $\ref{Fig5}$, we show as an example the schematic diagram of a
3-layer cross structure, which consists of three alternating layers
of gold crosses and polymer separation layers, and a gold film at
the bottom. The geometric parameters of each cross/ polymer layer
are marked in Fig. $\ref{Fig5}$. The crosses from the bottom to the
top are denoted in sequence as the $1^{st}$ cross, $2^{nd}$ cross,
$3^{rd}$ cross. In the present design for 1-, 2-, or 3-layer
structure, wire width $w$ is equal (for simplicity) and fixed to $w
= 6$ ${\rm {\mu}m}$ for all the crosses. The other geometrical
parameters for different layers of the structure are optimized and
given in Table $\ref{tab:table1}$. The corresponding absorption
spectra for normally incident plane wave with electric field in x
direction (see Fig. $\ref{Fig5}$) are simulated and presented in
Fig. $\ref{Fig6}$(a), (b), (c). For the sake of clarity, we first
study the case of 2-layer structure. Different from the case of the
1-layer structure with single resonance (I$_a$), two closely
positioned resonances with absorption up to $99.9\%$ are clearly
observed in the inset of Fig. $\ref{Fig6}$(b). One resonance
(II$_a$) is at frequency $\omega = 4.56$ ${\rm THz}$, while the
other resonance (II$_b$) is at frequency $\omega = 4.85$ ${\rm
THz}$. To understand the origin of the spectral characteristics, the
distributions of y-component magnetic field magnitude $|H_y|$ in
plane $y = 0$ (see Fig. $\ref{Fig5}$) at the two resonances are
shown in Fig. $\ref{Fig7}$(b) and (c). As Fig. $\ref{Fig7}$(b)
shows, resonance II$_a$ of lower frequency is primarily associated
with the excitation of magnetic polariton in the first layer, which
is caused by the electric dipole coupling between the 1st cross and
the metallic film [like the case of 1-layer structure, see Fig.
$\ref{Fig7}$(a)], while only a small magnetic response can be
observed in the second layer at this frequency. For resonance
II$_b$, strong magnetic field is found in both the first and second
layers as shown in Fig. $\ref{Fig7}$(b), which means that the
magnetic polariton of each layer contributes significantly to this
resonant absorption. Thus resonance II$_b$ can be regarded as a
hybridized mode \cite{liu1,liu2} of two magnetic polaritons strongly
coupled to each other. Due to the great contribution from the
magnetic polariton of the second layer, resonance II$_b$ shifts
slightly to a higher frequency as compared with resonance II$_a$.
Owing to these two closely positioned resonant peaks, we obtain a
relatively wide frequency band (from 4.45 ${\rm THz}$ to 4.95 ${\rm
THz}$) of absorption, where nearly perfect absorption (more than
$97\%$) occurs. The full bandwidth at half maximum (FWHM) of the
absorption is greatly improved to $27\%$ (with respect to the
central frequency), which is almost two times larger than that of
the 1-layer structure. Taking a step further, we demonstrate a
broader bandwidth absorption in a three-layer structure in Fig.
$\ref{Fig6}$(c). By stacking one more layer, additional magnetic
polariton is introduced to this hybridization system. Accordingly,
three closely located resonances (III$_a$, III$_b$, III$_c$) are
observed at frequencies $\omega = 4.55 $ ${\rm THz}$, 4.96 ${\rm
THz}$ and 5.37 ${\rm THz}$, as shown in the inset of Fig.
$\ref{Fig6}$(c). The corresponding distributions of y-component
magnetic field magnitude $|H_y|$ on plane $y = 0$ are also shown in
Fig. $\ref{Fig7}$(d), (e), (f). Each resonance is a hybridized mode
contributed differently by the three magnetic polaritons of the
three layers. In this 3-layer cross structure, a $1.03$ ${\rm THz}$
frequency band ($4.44 \sim 5.47$ ${\rm THz}$) with nearly perfect
absorption (more than $97\%$) is achieved, and the FWHM of the
absorption increases to 1.9 ${\rm THz}$, which is nearly $38\%$ of
the central frequency. Meanwhile, the thickness of the 3-layer
structure is still quit thin (no more than $\lambda/15$ ).
Importantly, the wide-angle feature of this high absorption caused
by the magnetic polariton is preserved in this multi-layer structure
as shown in Fig. $\ref{Fig8}$(a), (b) for TE and TM waves,
respectively. The absorption variation with the change of the
azimuthal angle is not given here, as the stacking of layers in
z-direction does not spoil the structural symmetry in the x-y plane
[consequently the isotropic absorption in the x-y plane can be
obtained just like the case of single layer, see Fig.
$\ref{Fig4}$(b), (d)]. A further broadening of the absorption
bandwidth is also possible by increasing the number of stacked
layers.

\begin{table*}
\caption{\label{tab:table1}Optimized parameters (${\rm {\mu}m}$) for
1-, 2-, and 3-layer cross structures }
\begin{ruledtabular}
\begin{tabular}{cccccccc}
\qquad  &$l_1$& $l_2$ & $l_3$ & $t_1$ & $t_2$ & $t_3$ & $w$\\
\hline
1-layer & 17  & $-$   & $-$  & 1.6  & $-$   & $-$  & 6\\
2-layer & 17  & 16.5  & $-$  & 0.9  & 2.2  & $-$  & 6  \\
3-layer & 17  & 15.4  & 15  & 0.7  & 1.1  & 2.0  & 6  \\
\end{tabular}
\end{ruledtabular}
\end{table*}

\section{Conclusion}
In summary, we have designed a thin THz absorber with a thickness
smaller than $\lambda/25$, and achieved a high absorption which is
nearly omni-directional for both TE and TM polarizations. A
simplified LC model has been introduced to elucidate the resonant
behavior. The wide angle nature of the resonant absorption has been
carefully analyzed. Moreover, we have successfully demonstrated that
the bandwidth of the absorption can be effective improved by using
multi-layer structure, while the wide angle feature remains. The
proposed multi-layer structure can be easily fabricated through a
layer-by-layer technique \cite{Qi,Subramania,Chang}, and allows for
easy integration with various devices. This omni-directional THz
absorber, which is also polarization-insensitive and broad band, has
potential applications in e.g. thermal detectors and THz imaging.

\begin{acknowledgments}
The authors would like to acknowledge the partial support of
National Basic Research program (973 Program) of China under Project
No.2004CB719800 and a Swedish Research Council (VR) grant on
metamaterials.
\end{acknowledgments}


\end{document}